# Diversity and Interdisciplinarity:

# How Can One Distinguish and Recombine Disparity, Variety, and Balance?

Loet Leydesdorff *

**Abstract**

The dilemma which remained unsolved using Rao-Stirling diversity, namely of how variety and balance can be combined into "dual concept diversity" (Stirling, 1998, pp. 48f.) can be clarified by using Nijssen *et al.*'s (1998) argument that the Gini coefficient is a perfect indicator of balance. However, the Gini coefficient is not an indicator of variety; this latter term can be operationalized independently as relative variety. The three components of diversity—variety, balance, and disparity—can thus be clearly distinguished and independently operationalized as measures varying between zero and one. The new diversity indicator ranges with more resolving power in the empirical case.

 *corresponding author; Amsterdam School of Communication Research (ASCoR), University of Amsterdam
PO Box 15793, 1001 NG Amsterdam, The Netherlands; loet@leydesdorff.net

1. **Introduction**

Rao-Stirling diversity is increasingly used as a measure of interdisciplinarity in bibliometrics (e.g., Rafols & Meyer, 2010; Leydesdorff, Kogler, & Yan, 2017; cf. Zhou *et al.*, 2012). In a brief communication entitled "The Repeat Rate: From Hirschman to Stirling," Ronald Rousseau argues that—"contrary to a recent statement in Leydesdorff *et al.* (2018, p. 573)" —that this index (Rao, 1982) or its monotone transformations (Zhang *et al.*, 2016) includes the *three* aspects of variety, balance, and disparity as distinguished, for example, by Stirling (2007) and Rafols & Meyer (2010). Rao-Stirling diversity, however, is defined in terms of two factors, as follows:

$$\Delta = \sum_{\substack{i,j=1 \\ i \neq j}}^{n} (p_i p_j)(d_{ij}) \tag{1}$$

where $d_{ij}$ is a disparity measure between two classes $i$ and $j$, and $p_i$ is the proportion of elements assigned to each class $i$.

I added the brackets in Eq. 1 to show that Rao-Stirling diversity is composed of two factors: The right-hand factor operationalizes disparity; the left-hand one is also known as the Hirschman-Herfindahl or Simpson index.[2] It seems to me that two factors cannot cover three concepts unless one uses two words for the same operationalization. However, Rousseau (2018) argues that the left-hand term of Eq. 1 measures both variety and balance.

---

[2] $\sum_{ij} p_i p_j = 1$ when taken over all *i* and *j*. The Simpson index is equal to $\Sigma_i (p_i)^2$, and the Gini-Simpson to $[1 - \Sigma_i (p_i)^2]$.

Rousseau *et al.* (1999) already addressed the issue when they formulated as follows (at p. 213):

> It is generally agreed that diversity combines two aspects: species richness and evenness. Disagreement arises at how these two aspects should be combined, and how to measure this combination, which is then called "diversity".

How and why are these two aspects of diversity compared and integrated in the left-hand term of Eq. 1? Following Junge (1994), Stirling (1998, at p. 48) suggests labeling this integration as "dual concept diversity" and notes that "to many authorities in ecology, dual concept diversity is synonymous with diversity itself."

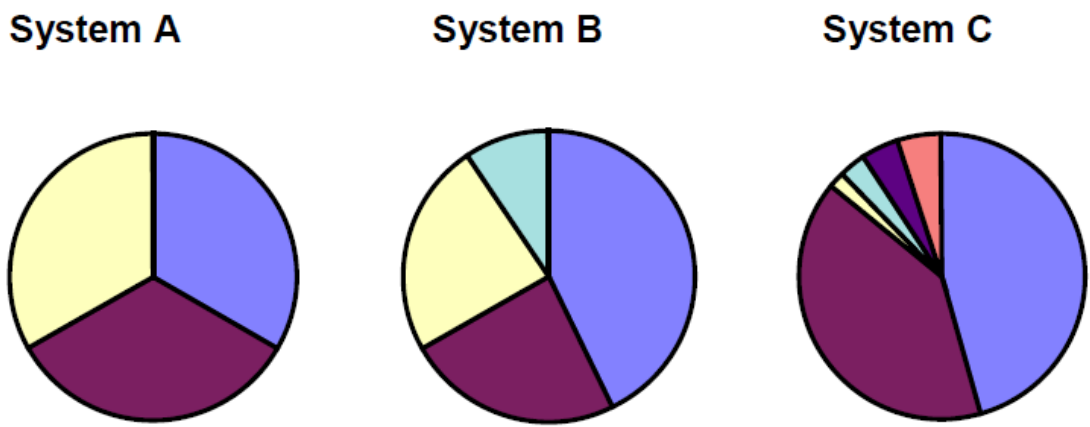

**Figure 1**: The question of the relative priority assigned to variety and balance in dual concept diversity. Source: Stirling (1998, at p. 49).

Using Figure 1, Stirling (1998) shows the possible dilemma when combining the two "subordinate properties" into a single "dual concept" when he formulates as follows at p. 48:

Where variety is held to be the most important property, System C might reasonably be held to be most (dual concept) diverse. Where a greater priority is attached to the evenness in the balance between options, System A might be ranked highest. In addition, there are a multitude of possible intermediate possibilities, such as System B.

Stirling (1998) then discusses at length the possibility to use the Simpson index or Shannon-diversity for the measurement of "dual concept diversity" and concludes (on p. 57) that '***there are good reasons to prefer the Shannon function as a robust general "non-parametric" measure of dual concept diversity***' (boldface and italics in the original.) Nevertheless, the Simpson index is most frequently used in the literature for this purpose (Stirling, 2007).[3]

## 2. An alternative operationalization of diversity

In a study of the Lorenz curve as a graphical representation of "evenness" or "balance," Nijssen, Rousseau, & Van Hecke (1998) proved that both the Gini index and the coefficient of variation (that is, the standard deviation divided by the mean of the distribution or, in formula format, $\sigma/\mu$) are perfect indicators of balance (Rousseau, personal communication, 16 March 2018). (The coefficient of variation is not bounded between zero and one.) Additionally, the Gini index is *not* a measure of variety (Rousseau, in press, p. 6).

*Variety* is the number of categories into which system elements are apportioned (Stirling, 2007, p. 709), for example, the number of species ($N$) in an eco-system (MacArthur, 1965). The problem with integrating this measure into an index of diversity might be that *N* is not bound between zero and one. I suggest solving this by using *n/N*, that is, the relative variety: *n* denotes the number of categories with values larger than zero, whereas *N* denotes the number of possible categories. In the example which I will elaborate below, for example, among the 654 classes for patents in the so-called CPC classification at the USPTO, Amsterdam's portfolio shows a value in 131 of them: the relative variety *n/N* is therefore 131/654 = 0.20.

---

[3] Hill (1973) derived that the two indicators can be considered as variants of a general formalization. See Stirling (1998, at pp. 49f) for the elaboration.

In the discussion about related and unrelated variety, Frenken *et al.* (2007) proposed Shannon entropy as a measure of "unrelated variety." As a measure of "related variety" these authors use Theil's (1972) decomposition algorithm for appreciating the grouping (cf. Leydesdorff, 1991). However, this measure assumes the *ex ante* definition of relevant groups. The disparity matrix operates in terms of ecological distances and is not based on such *a priori* assumptions about structure (Izsák & Papp, 1995). Relatedness is already covered by the term $d_{ij}$ in Eq. 1. Shannon entropy can be normalized relative to the maximum entropy and then varies between zero and one (or as percentage entropy). If one wishes to appreciate not only the number of categories but also the values, Shannon entropy could be an alternative for measuring variety. Grouping is not advised, because the disparity measure already covers the ecological distances that can indicate relatedness.

## 3. An empirical elaboration

If one wishes to consider the three aspects of diversity—variety, balance, and disparity—in a single measure equivalent to Rao-Stirling diversity, one thus can multiply the corresponding elements in the disparity matrix with the values of the Gini index and relative variety. All three factors are bounded between zero and one and are decomposable. (Note that the coefficient of variation is not bound between zero and one.) One thus obtains the following diversity measure for each city *c*:

$$Div_c = (n_c/N) * Gini_c * [\sum_{\substack{i=1, \\ j=1, \\ i \neq j}}^{\substack{i=n_c \\ j=n_c}} d_{ij} / \{n_c * (n_c - 1)\}]$$

The first term is the relative variety as defined above: the number of valued categories for this city (excluding zeros) divided by the total number of categories (that is in this case, 654; including zeros). The second term is the Gini coefficient of the vector of these $n_c$ categories, and the third weights the disparity as a measure for each observation permutating the cells *i* and *j* along the vector, but excluding the main diagonal.[4] The normalization in the third component is needed for warranting that the disparity values (e.g., the Euclidean distance or (1 – *cosine*)) function as weightings between zero and one. As in the case of Rao-Stirling diversity, the cosine-values are taken from the symmetrical cosine-matrix among the 654 column vectors of the asymmetrical matrix of 654 categories versus more than five million patents used by Leydesdorff *et al.* (2017).[5]

For the computation of the Gini coefficient, I follow Buchan's (2002) simplification of the computation which the author formulated as follows:

---

[4] If one wished, one could replace the variety measure with the Shannon function.
[5] A routine for the computation can be found at http://www.leydesdorff.net/software/diverse (see Appendix I).

The classical definition of G appears in the notation of the theory of relative mean difference:

$$G = \frac{\sum_{i=1}^{n} \sum_{j=1}^{n} |x_i - x_j|}{2n^2 \bar{x}} \quad (2)$$

where *x* is an observed value, *n* is the number of values observed and *x bar* is the mean value.

If the *x* values are first placed in ascending order, such that each *x* has rank *i*, some of the comparisons above can be avoided and computation is quicker:

$$G = \frac{2}{n^2 \bar{x}} \sum_{i=1}^{n} i(x_i - \bar{x})$$

$$G = \frac{\sum_{i=1}^{n} (2i - n - 1) x_i}{n \sum_{i=1}^{n} x_i}$$

where *x* is an observed value, *n* is the number of values observed and *i* is the rank of values in ascending order.

In the following example from Leydesdorff *et al*. (2017), disparity is measured as (1 – cosine) between each two distributions (Jaffe, 1989). In this study we compared 20 cities (four cities each in five countries) in terms of the Rao-Stirling diversity of their patent portfolios operationalized as patents granted by the USPTO in 2016. The results are provided in Table 5 (at p. 1584) of that study and compared here below in Table 1 with the values for the new indicator in the right-hand column.

**Table 1**: Rank-ordered list of twenty cities in terms of the diversity of patent portfolios granted at the USPTO in 2016. Source of the left-hand column: Leydesdorff *et al.* (2017, Table 5 at p. 1584).

| City | Rao | City | Diversity |
|---|---|---|---|
| Paris | 0.83 | Shanghai | 0.74 |
| Boston | 0.80 | Beijing | 0.71 |
| Rotterdam | 0.80 | Paris | 0.62 |
| Jerusalem | 0.79 | Atlanta | 0.61 |
| Atlanta | 0.78 | Boulder | 0.52 |
| Eindhoven | 0.78 | Boston | 0.49 |
| Nanjing | 0.78 | Berkeley | 0.45 |
| Berkeley | 0.78 | Telaviv | 0.42 |
| Shanghai | 0.78 | Eindhoven | 0.41 |
| Boulder | 0.78 | Haifa | 0.36 |
| Beersheva | 0.78 | Grenoble | 0.33 |
| Amsterdam | 0.76 | Jerusalem | 0.29 |
| Beijing | 0.71 | Toulouse | 0.27 |
| Toulouse | 0.71 | Amsterdam | 0.25 |
| Telaviv | 0.71 | Nanjing | 0.23 |
| Marseille | 0.70 | Rotterdam | 0.15 |
| Haifa | 0.69 | Beersheva | 0.12 |
| Grenoble | 0.69 | Dalian | 0.10 |
| Dalian | 0.69 | Wageningen | 0.09 |
| Wageningen | 0.50 | Marseille | 0.03 |

Whereas the left-hand ranking is counter-intuitive in placing Rotterdam and Jerusalem above, for example, Shanghai and Beijing, these latter two cities are attributed the highest rankings using the new indicator. Furthermore, the Rao-Stirling diversity ranges from 0.50 (Wageningen) to 0.83 (Paris), whereas the new diversity index ranges from 0.03 (Marseille) to 0.74 (Shanghai). Figure 2 shows these ranges graphically. The new diversity measure has a stronger resolving power than Rao-Stirling diversity.

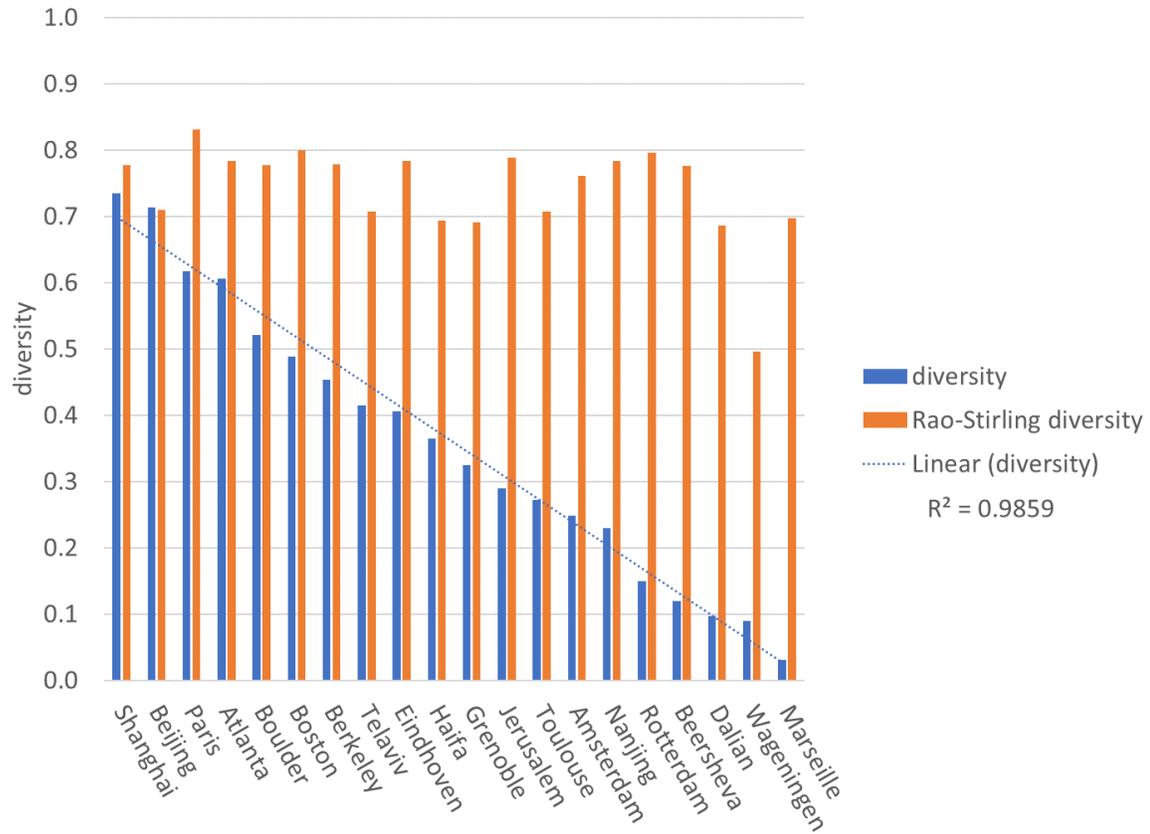

**Figure 2**: Rao-Stirling diversity and the diversity measure proposed here for the patent portfolios of twenty cities in terms of the CPC classification at the USPTO for patents granted in 2016.

The cities under study were chosen so that one could expect differences among them; however, these were smaller than expected using Rao-Stirling diversity. For example, Boston and Rotterdam had the same value on this indicator. Using the new diversity measure, however, the diversity of the portfolio of Boston is more than three times higher than that of Rotterdam.

Table 3: Pearson correlation coefficients in the lower triangle and Spearman's rank-order correlations in the upper triangle.

|  | Rao-Stirling | Diversity | Gini | Variety | Simpson | Shannon |
|---|---|---|---|---|---|---|
| *Rao-Stirling* |  | 0.438 | -0.084 | 0.470* | 0.874** | 0.893** |
| *Diversity* | 0.417 |  | 0.747** | 0.997** | 0.416 | 0.589** |
| *Gini* | -0.078 | 0.765** |  | 0.721** | -0.092 | 0.060 |
| *Variety* | 0.492* | 0.992** | 0.714** |  | 0.443 | 0.623** |
| *Simpson* | 0.896** | 0.346 | -0.114 | 0.412 |  | 0.925** |
| *Shannon* | 0.890** | 0.600** | 0.184 | 0.684** | 0.835** |  |

\*\* Correlation is significant at the 0.01 level (2-tailed).
 \* Correlation is significant at the 0.05 level (2-tailed).

Table 3 provides the relevant correlations: Spearman rank-order correlations are shown in the upper triangle and Pearson correlations on the basis of comparing among these twenty cities in the lower triangle. As could be expected, Rao-Stirling diversity correlates with the Simpson index and Shannon diversity, but not with the Gini coefficient.[6] The new diversity measure is *not* significantly correlated with Rao-Stirling diversity or the Simpson index, but—not surprisingly—with the Gini coefficient and with variety; these two factors are constitutive for the diversity in this approach in addition to the disparity.

**Conclusions and discussion**

The dilemma which remained unsolved using Rao-Stirling diversity, namely of how variety and balance can be combined into "dual concept diversity" (Stirling, 1998, pp. 48f.), can be clarified using Nijssen *et al.*'s (1998) argument that the Gini coefficient is a perfect indicator of balance.

---

[6] As can be expected, the coefficient of variation correlated significantly with the Gini coefficient: both Spearman's rank-order correlation and the Pearson correlation are .94 ($p < .01$; $n = 20$).

Since the Gini coefficient is not an indicator of variety; this latter term can be operationalized as relative variety and thus be bounded between zero and one. The three components of diversity—variety, balance, and disparity—can thus be clearly distinguished and independently operationalized as measures varying between zero and one. The new diversity indicator ranges with more resolving power in the empirical case. However, the new diversity indicator did not correlate with Rao-Stirling diversity.

I don't want to argue for this diversity measure beyond the status of another indicator. Unlike the confusion hitherto, however, the new indicator is based on the solution made possible by Nijssen *et al.*'s (1998) proof and Stirling's (1998) analysis of the literature. The independent operationalization of the three aspects of diversity distinguished by Stirling (1998, 2007) provides a more reliable ground than "dual" or higher-order concepts. A routine is provided at http://www.leydesdorff.net/software/diverse for computing both Rao-Stirling diversity and this new indicator (Appendix I).

The diversity issue is important for the measurement of interdisciplinarity and knowledge integration in science and technology studies. However, the further elaboration of this relevance requires yet another discussion (e.g., Wagner *et al*., 2011). In Leydesdorff *et al.* (2018), for example, we argued that a high diversity—measured as Rao-Stirling diversity—in citing patterns may indicate esoteric originality at the journal level and perhaps trans-disciplinarity more than knowledge integration. Uzzi *et al*. (2013), however, considered atypical combinations in citing behavior at the paper level on the contrary as an indication of novelty.

**Acknowledgment**

I thank Ronald Rousseau for comments and stimulating discussions about previous versions of this communication.

Stirling, A. (1998). On the economics and analysis of diversity. *SPRU Electronic Working Paper Series* No. 28, at
   http://www.sussex.ac.uk/Units/spru/publications/imprint/sewps/sewp28/sewp28.pdf.
Stirling, A. (2007). A general framework for analysing diversity in science, technology and society. *Journal of the Royal Society Interface, 4*(15), 707-719.
Theil, H. (1972). *Statistical Decomposition Analysis*. Amsterdam/ London: North-Holland.
Uzzi, B., Mukherjee, S., Stringer, M., & Jones, B. (2013). Atypical combinations and scientific impact. *Science, 342*(6157), 468-472.
Wagner, C. S., Roessner, J. D., Bobb, K., Klein, J. T., Boyack, K. W., Keyton, J., . . . Börner, K. (2011). Approaches to Understanding and Measuring Interdisciplinary Scientific Research (IDR): A Review of the Literature. *Journal of Informetrics, 5*(1), 14-26.
Zhang, L., Rousseau, R., & Glänzel, W. (2016). Diversity of references as an indicator for interdisciplinarity of journals: Taking similarity between subject fields into account. *Journal of the Association for Information Science and Technology, 67*(5), 1257-1265. doi: 10.1002/asi.23487
Zhou, Q., Rousseau, R., Yang, L., Yue, T., & Yang, G. (2012). A general framework for describing diversity within systems and similarity between systems with applications in informetrics. *Scientometrics, 93*(3), 787-812.
13

**Appendix I**

The program div.exe can be retrieved at http://www.leydesdorff.net/software/diverse .

**Input files** are:

- Matrix.csv contains the data to be analyzed. Div.exe analyzes column vectors. The file needs to be in .csv (comma-separated variable) style and saved as MS-DOS. The file should not contain a header with variable labels, but only numerical information.
  For example:

  ```
  0,2,0,0,0
  2,1,0,0,5
  0,0,0,0,0
  0,0,0,0,0
  27,0,0,27
  0,0,0,0,0
  0,0,0,0,0
  0,0,0,0,0
  0,0,0,0,0
  0,0,8,5,0
  ```

  In the case under study in this paper, twenty cities are compared in terms of 654 classes of patents. The matrix has twenty columns and 654 rows.

- Sim.csv contains a symmetrical similarity matrix (e.g., cosine values) in csv-format without a header.
  For example:

  ```
  1.0000,0.6270,0.3146,0.1280,0.1564
  0.6270,1.0000,0.1319,0.0777,0.2190
  0.3146,0.1319,1.0000,0.4214,0.1322
  0.1280,0.0777,0.4214,1.0000,0.0865
  0.1564,0.2190,0.1322,0.0865,1.0000
  ```

  In the case under study in this paper, the comparison is in terms of 654 classes. The cosine matrix is a symmetrical (1-mode) matrix of 654 * 654 cells with ones on the main diagonal. This file can be retrieved at https://www.leydesdorff.net/cpc_cos/portfolio/cos_cpc.dbf. (Save the file from https://www.leydesdorff.net/cpc_cos/portfolio/ using the right-side mouse knob.)

The output file **diverse.dbf** contains the following information for each vector:

- The first column contains the number of the column vector of matrix.csv analyzed.
- Rao-Stirling diversity;
- **Diversity** as defined in this study;
- Gini;
- Simpson;

- Shannon;
- $H_{max}$
- Variety;
- Total number of cases;
- Number of cases with a value larger than one.